\documentclass[12pt]{article}

\usepackage{amsmath}
\usepackage{graphicx,psfrag,epsf}
\usepackage{enumerate}
\usepackage{natbib}
\usepackage{url} 
\usepackage{enumitem}
\usepackage{subcaption} 
\usepackage{mathtools}
\usepackage{booktabs}
\usepackage{multirow}
\usepackage{array}
\usepackage{algorithm}
\usepackage{algpseudocode}
\usepackage{float} 
\usepackage{color}
\usepackage{amsmath}
\usepackage{amssymb}
\usepackage{amsthm}

\usepackage[colorlinks=true,
            linkcolor=blue,    
            citecolor=blue,    
            urlcolor=blue,     
            anchorcolor=blue]{hyperref}

\newcommand{\blind}{1}

\def \btheta {\boldsymbol{\theta}}
\def \be {\boldsymbol{e}}
\def \bg {\boldsymbol{g}}

\def \R {\mathbb{R}}

\def \s {\mathbf{s}}

\DeclareMathOperator*{\argmin}{argmin}

\begin{document}

\def\spacingset#1{\renewcommand{\baselinestretch}%
{#1}\small\normalsize} \spacingset{1}

\newtheorem{assumption}{Assumption}
\newtheorem{proposition}{Proposition}
\newtheorem{theorem}{Theorem}
\newtheorem{remark}{Remark}
\newtheorem{lemma}{Lemma}
\newtheorem{corollary}{Corollary}
\newtheorem{definition}{Definition}
\newtheorem{example}{Example}

\if1\blind
{
  \title{\bf Preference-based Centrality and Ranking in General Metric Spaces}
  \author{Lingfeng Lyu\textsuperscript{1,2},    Doudou Zhou\textsuperscript{1}\thanks{Corresponding author: ddzhou@nus.edu.sg}\bigskip \\
    \small 
    \textsuperscript{1} Department of Statistics and Data Science, National University of Singapore, Singapore \\
        \small 
    \textsuperscript{2} Department of Statistics and Finance, University of Science and Technology of China}
    \date{ }
  \maketitle
} \fi

\if0\blind
{
  \bigskip
  \bigskip
  \bigskip
  \begin{center}
    {\LARGE\bf Preference-based Centrality and Ranking in General Metric Spaces}
\end{center}
  \medskip
} \fi

\bigskip

\begin{abstract}
Ranking or assessing centrality in multivariate and non-Euclidean data is difficult because there is no canonical order and many depth notions become computationally fragile in high-dimensional or structured settings. We introduce a preference-based notion of centrality defined through population proximity comparisons with respect to a random reference draw, yielding a metric-intrinsic statistical functional that is well-defined on general metric spaces. Because the induced pairwise preferences may be non-transitive, we map them to a coherent one-dimensional score via a Bradley--Terry--Luce cross-entropy projection, viewed as a calibrated aggregation device rather than a correctly specified model. We develop two finite-sample estimators a convex M-estimator and a fast spectral estimator based on a comparison operator, and establish identifiability and consistency under mild conditions. Simulations and real-data examples, including high-dimensional and functional observations, illustrate that the proposed scores provide stable, interpretable rankings aligned with the underlying preference centrality.
\end{abstract}


\vfill

\newpage
\spacingset{1.45} 
\section{Introduction}
\label{sec:intro}

Centrality is a fundamental concept in data analysis. It underlies representative point selection, outlier detection, risk stratification, and the construction of interpretable orderings. In univariate settings, the real line provides a canonical order, so centrality and ranking are naturally defined through ranks, quantiles, and order statistics. In contrast, modern data are frequently multivariate or live on non-Euclidean spaces such as manifolds, networks, graphs, and learned embedding spaces, where no intrinsic notion of order is available. In these settings, defining population-level centrality and estimating a stable, interpretable ordering from finite samples remain challenging problems. In many contemporary applications, the only primitive is a dissimilarity or distance oracle (possibly task-specific or learned), and there may be no meaningful linear, convex, or coordinate structure to exploit.

A broad literature has sought to generalize ranking and centrality beyond the univariate case.
Classical approaches include componentwise and spatial ranks \citep{Bickel1965,hallin1995multivariate,puri1966class,chaudhuri1996geometric,oja2010multivariate}, depth-induced orderings \citep{liu1993quality,zuo2000general}, Mahalanobis and metric-based constructions \citep{hallin2002optimal,hallin2006parametric,pan2018ball}, and more recent transport- or graph-based formulations \citep{deb2021multivariate,zhou2023new}.
Statistical depth, in particular, has emerged as a principled framework for multivariate centrality, formalizing center--outward ordering through axioms such as affine equivariance, maximality, monotonicity, and robustness \citep{Tukey1975,Liu1990,VardiZhang2000,Zuo2003,Parzen1962}. Depth notions have also been developed for functional data and come with accompanying theoretical guarantees \citep{Yeon2025JRSSB, Bocinec2025projection}. 
Despite their success, many existing depth notions rely on linear or convex structure and can lose resolution or become computationally burdensome in high dimensions or in complex geometries, where distance concentration, hubness, and anisotropy obscure purely geometric notions of centrality \citep{Beyer1999,Radovanovic2010,Feldbauer2019}.
More broadly, when only a general dissimilarity is available, it is often unclear what population object a given surrogate score targets, and how pairwise information across the sample should be coherently aggregated into a global ordering.

This paper takes a different starting point.
Rather than constructing centrality directly from global geometric summaries, we ground it in \emph{reference-based pairwise comparisons}.
Given two candidates, we ask which one tends to be closer to a typical draw from the underlying distribution.
Such comparisons are intrinsic, require only a dissimilarity measure, and remain well-defined on arbitrary metric spaces. 
Formally, let $F$ be a probability measure on a space $\mathcal X$ equipped with a measurable dissimilarity function $\delta:\mathcal X\times\mathcal X\to[0,\infty)$.
We define the population preference
\[
p_F(x,y) = \mathbb P_{Z\sim F}\{\delta(Z,x) > \delta(Z,y)\},
\]
and the induced centrality functional
\begin{equation}
r_F(y) = \mathbb E_{X\sim F}[p_F(X,y)].
\label{eq:r-def}
\end{equation}
Throughout, we focus on the no-ties regime where $p_F(x,y)+p_F(y,x)=1$ for all distinct $x,y\in\mathcal X$.
Intuitively, $r_F(y)$ is the probability that $y$ ``wins'' against a random opponent when judged by a random reference point; it coincides with the population counterpart of an average win-rate aggregation in paired-comparison settings. This functional defines a population notion of centrality determined solely by $(F,\delta)$ and is directly meaningful in Euclidean, high-dimensional, and non-Euclidean settings.

Estimating and using $r_F(\cdot)$ in practice, however, raises two fundamental challenges. First, the induced pairwise preferences need not be transitive: it is possible to have $x$ preferred to $y$, $y$ preferred to $z$, and $z$ preferred to $x$, even at the population level. Second, empirical estimates of $p_F(x,y)$ are noisy and strongly dependent, as they are constructed from shared reference points. As a result, naive aggregation schemes, such as empirical win rates, can yield unstable rankings and do not provide a calibrated score scale with a likelihood-based inferential handle.

To address these issues, we aggregate induced preferences through a calibrated one-dimensional projection.
Specifically, we consider a Bradley--Terry--Luce (BTL) type representation \citep{bradley1952rank,luce1959individual} that maps pairwise preferences onto a transitive score scale by minimizing a population cross-entropy risk.
Importantly, this formulation is used as a \emph{projection} rather than as a correctly specified model: the resulting score function is defined as the unique minimizer of a well-posed M-estimation problem and exists regardless of whether the BTL form exactly matches the underlying preference probabilities.
This projection enforces global coherence of the ordering, regularizes noisy comparisons, and yields interpretable log-odds differences that provide a coherent and computationally tractable score scale.

We call the resulting framework CORE (Centrality Ordering via Relative Evaluation). At the population level, we show that the calibrated score is a strictly monotone transformation of $r_F(\cdot)$, ensuring that the projection preserves the intrinsic preference-based ordering while providing a stable coordinate system. At the sample level, we develop two complementary estimators: a convex optimization-based estimator and a fast spectral estimator based on the stationary distribution of an empirical comparison operator. We establish identifiability, consistency, and stability of the proposed scores under mild conditions. In Euclidean settings where a center--outward geometry is meaningful, we further show that, under standard regularity conditions, the induced population score satisfies the canonical axioms associated with statistical depth. Through simulations and real-data analyses, we illustrate that CORE produces stable rankings aligned with the underlying preference centrality across high-dimensional and non-Euclidean settings.

The remainder of the paper is organized as follows.
Section~\ref{sec:core} introduces the CORE framework and the proposed estimators.
Section~\ref{sec:theory} develops the theoretical properties of the population and sample-level scores.
Section~\ref{sec:sim} presents simulation studies and real data analysis is performed in Section~\ref{sec:real}.
Section~\ref{sec:conclusion} concludes with a discussion and directions for future work.

\section{Method}
\label{sec:core}

\subsection{From induced preferences to a calibrated centrality score}

The population preference function $p_F(\cdot,\cdot)$ encodes reference-based pairwise comparisons induced by $(F,\delta)$. While these comparisons are intrinsic and well-defined on general dissimilarity spaces, $p_F$ is a bivariate object and does not directly yield a scalar notion of centrality or a coherent global ordering on $\mathcal X$. In particular, the induced majority relation $x \succ y$ defined by $p_F(y,x)>1/2$ need not be transitive and may exhibit preference cycles, even at the population level; see Supplementary~S.1 for an explicit example. These features motivate aggregating pairwise preferences through a calibrated one-dimensional representation that enforces global coherence while retaining the intrinsic comparison structure.

We adopt a BTL-type formulation as a projection device. Let $\theta:\mathcal X \to \mathbb R$ be a real-valued score function and define
\begin{equation}
\label{eq:BTL}
q_\theta(x,y)
=
\frac{\exp\{\theta(y)\}}{\exp\{\theta(x)\}+\exp\{\theta(y)\}},
\qquad x,y\in\mathcal X.
\end{equation}
Larger values of $\theta(\cdot)$ correspond to higher centrality, and the induced ordering is transitive by construction. Importantly, we do not assume that $q_\theta$ correctly specifies the true preference probabilities $p_F$. Instead, the BTL form is used as a calibrated projection that maps possibly cyclic preferences onto a coherent score scale.

We formalize this projection through a population M-estimation problem. Since $p_F(x,y)\in(0,1)$ represents a comparison probability, a natural way to project possibly non-transitive preferences onto a parametric comparison scale is via a strictly proper scoring rule. We therefore quantify the discrepancy between $p_F$ and $q_\theta$ using the cross-entropy risk
\begin{equation}
\label{eq:population-risk-theta}
\mathcal L_F(\theta)
=
\mathbb E_{X,Y\sim F}\left[
-p_F(X,Y)\log q_\theta(X,Y)
-\{1-p_F(X,Y)\}\log\{1-q_\theta(X,Y)\}
\right].
\end{equation}
Up to an additive term that does not depend on $\theta$, minimizing $\mathcal L_F(\theta)$ is equivalent to minimizing the expected Kullback--Leibler divergence from the Bernoulli distribution with success probability $p_F(X,Y)$ to its logistic approximation $q_\theta(X,Y)$, averaged over $(X,Y)\sim F$. This defines a well-posed projection of the preference kernel onto the logistic comparison family.

Since $q_\theta(x,y)$ depends only on score differences $\theta(y)-\theta(x)$, the risk $\mathcal L_F(\theta)$ is invariant under additive shifts of $\theta$. To ensure identifiability, we work with centered score functions.
Let $L^2(F)$ denote the space of measurable functions $\theta:\mathcal X\to\mathbb R$ with $\int \theta(x)^2\,dF(x)<\infty$, and let $L^\infty(F)$ denote the space of essentially bounded measurable functions. Define
\[
L^2_0(F)=\{\theta\in L^2(F):\mathbb E_F[\theta(X)]=0\}.
\]
Let $\Theta(F)\subset L^2(F)$ be a user-specified function class. In the most general case one may take $\Theta(F)=L^2(F)\cap L^\infty(F)$ to ensure square integrability and boundedness; more structured classes (e.g., RKHS or parametric families) can also be considered. We define $\Theta_0(F)=\Theta(F)\cap L^2_0(F)$ and set
\begin{equation}
\label{eq:theta-star-def}
\theta^\star \in \argmin_{\theta\in\Theta_0(F)} \mathcal L_F(\theta).
\end{equation}
The resulting $\theta^\star$ is a population-level calibrated centrality score. As shown in Section~\ref{sec:theory}, $\theta^\star(\cdot)$ preserves the intrinsic ranking induced by $r_F(\cdot)$ up to a strictly monotone transformation. For convenience, we also define the associated strength function $s^\star(x)=\exp\{\theta^\star(x)\}$, which induces the same ordering as $\theta^\star$.

\subsection{Sample-level M-estimation of centrality scores}
\label{subsec:core_gd}

Let $x_1,\ldots,x_n$ be an i.i.d.\ sample from $F$. 
Recall that the population target is the function $\theta^\star(\cdot)$ defined through the M-projection in \eqref{eq:theta-star-def}. 
In practice, centrality is used to rank the observed sample points. 
Accordingly, rather than estimating $\theta^\star(\cdot)$ over the entire space $\mathcal X$, we approximate it by estimating its values at the observed data points, leading to a finite-dimensional score vector $\btheta=(\theta_1,\ldots,\theta_n)^\top$, where $\theta_i$ is intended to estimate $\theta^\star(x_i)$.

To construct an estimator consistent with the population projection, we replace the population risk $\mathcal L_F(\theta)$ by its empirical counterpart. 
Under the BTL representation \eqref{eq:BTL}, the comparison probability between $x_i$ and $x_j$ is $q_\theta(x_i,x_j)=\sigma(\theta_j-\theta_i)$, where $\sigma(t)=(1+e^{-t})^{-1}$. 
Let $\widehat{\mathbf P}=\{\hat p_{ij}\}_{i,j=1}^n$ denote an empirical estimate of the preference kernel $p_F(x_i,x_j)$. 
The sample objective is then defined as
\begin{equation}
\label{eq:loss-no-ridge}
\mathcal L_n(\btheta;\widehat{\mathbf P})
=
\sum_{1\le i<j\le n}
\Big[
-\hat p_{ij}\log \sigma(\theta_j-\theta_i)
-(1-\hat p_{ij})\log\{1-\sigma(\theta_j-\theta_i)\}
\Big],
\end{equation}
which is the empirical analogue of \eqref{eq:population-risk-theta}. 

As in the population case, the empirical objective depends only on score differences and is therefore invariant under global shifts of $\btheta$. 
We impose the centering constraint $\mathbf 1^\top\btheta=0$ to ensure identifiability and define the M-estimator
\begin{equation}
\label{eq:emp-problem}
\widehat\btheta
=
\argmin_{\btheta\in\Theta_n}
\mathcal L_n(\btheta;\widehat{\mathbf P}), 
\qquad 
\Theta_n:=\{\btheta\in\mathbb R^n:\mathbf 1^\top\btheta=0\}.
\end{equation}
The objective is convex in $\btheta$, and strictly convex on $\Theta_n$ under mild connectivity conditions on the comparison graph induced by $\widehat{\mathbf P}$, ensuring uniqueness of the solution. 
We compute $\widehat\btheta$ using projected first-order optimization; Algorithm~\ref{al:GD-CORE} summarizes the procedure. For faster optimization of \eqref{eq:emp-problem}, we also implement a quasi-Newton solver based on L-BFGS-B updates (Supplementary Section~S.2).

The estimated strength scores are $\widehat s_i=\exp(\widehat\theta_i)$, which approximate the population strengths $s^\star(x_i)=\exp\{\theta^\star(x_i)\}$. 
We define the sample preference center as 
\[
\widehat\mu = x_{\hat\imath}, 
\qquad 
\hat\imath\in\arg\max_{1\le i\le n}\widehat\theta_i.
\]

\paragraph{A fast spectral approximation.}
For large $n$, solving \eqref{eq:emp-problem} can be computationally demanding. 
As a scalable alternative, we construct a spectral approximation based on a comparison-induced Markov chain, closely related to classical spectral ranking methods for paired comparisons \citep{Rank2017}. 

Given the empirical preference matrix $\widehat{\mathbf P}=\{\hat p_{ij}\}_{i,j=1}^n$, define the transition matrix $\mathbf T=[T_{ij}]_{i,j=1}^n$ by
\[
T_{ij}
=
\frac{1}{n-1}\hat p_{ij}\,\mathbb I(i\neq j)
+
\Big(1-\frac{1}{n-1}\sum_{k\neq i}\hat p_{ik}\Big)\mathbb I(i=j).
\]
This construction defines a random walk on the sample points in which transitions favor preferred alternatives. 
Under mild irreducibility conditions on $\widehat{\mathbf P}$, $\mathbf T$ admits a unique stationary distribution $\widehat\s$ satisfying $\widehat\s=\mathbf T^\top\widehat\s$. 

We use $\widehat\s$ as a spectral strength score and define centered log-scores
\[
\widehat\theta_i
=
\log \widehat s_i
-
\frac{1}{n}\sum_{k=1}^n \log \widehat s_k,
\qquad i=1,\ldots,n,
\]
where $\widehat s_i$ denotes the $i$th entry of $\widehat\s$. 
This procedure replaces the global convex optimization in \eqref{eq:emp-problem} by an eigenvector computation while preserving the same preference aggregation structure. 
Algorithm~\ref{al:CORE-spectral} details the implementation.

\begin{algorithm}
\caption{CORE-GD: M-estimation via projected first-order optimization}
\label{al:GD-CORE}

\textbf{Input:}
pairwise preference matrix $\widehat{\mathbf P} = \{\hat p_{ij}\}_{i,j=1}^n$,
stepsize $\eta>0$,
tolerance $\varepsilon>0$,
maximum iteration $T_{\max}$.

\textbf{Initialize:} $\btheta^{(0)}\leftarrow \mathbf 0\in\mathbb R^n$.

\textbf{For $t=0,1,\ldots,T_{\max}-1$:} 
\begin{itemize}
  \item Compute the gradient of $\mathcal L_n$ at $\btheta^{(t)}$:
  \[
  \bg^{(t)}
  =
  \sum_{1 \le i<j \le n}
  \bigl(
    \sigma(\theta^{(t)}_j - \theta^{(t)}_i) - \hat p_{ij}
  \bigr)\,
  (\be_j - \be_i).
  \]
  \item Enforce the centering constraint by projecting the gradient:
  \[
  \bg^{(t)} \leftarrow \bg^{(t)} - \frac{\mathbf 1^\top\bg^{(t)}}{n}\,\mathbf 1 .
  \]
  \item Update:
  $\btheta^{(t+1)} \leftarrow \btheta^{(t)} - \eta\,\bg^{(t)}$.
  \item Stop if $\|\bg^{(t)}\|_\infty \le \varepsilon$.
\end{itemize}

\textbf{Output:} $\widehat{\btheta}=\btheta^{(t)}$. 
\end{algorithm}

\begin{algorithm}
\caption{CORE-Spectral: spectral scores via the stationary distribution}
\label{al:CORE-spectral}

\textbf{Input:}
pairwise preference matrix $\widehat{\mathbf P}=\{\hat p_{ij}\}_{i,j=1}^n$ with $\hat p_{ii}=0$,
tolerance $\varepsilon>0$,
maximum iterations $T_{\max}$.

\textbf{Step I (Transition matrix).}
Construct $\mathbf T=\{T_{ij}\}_{i,j=1}^n$ by
\[
T_{ij}
=
\frac{1}{n-1}\hat p_{ij}\,\mathbb I(i\neq j)
+
\Big(1-\frac{1}{n-1}\sum_{k\neq i}\hat p_{ik}\Big)\mathbb I(i=j).
\]

\textbf{Step II (Power iteration).}
Initialize $\s^{(0)}=\frac{1}{n}\mathbf 1$.
For $t=0,1,\ldots,T_{\max}-1$, iterate
\[
\s^{(t+1)}=\mathbf T^\top \s^{(t)}.
\]
Stop if $\|\s^{(t+1)}-\s^{(t)}\|_1\le \varepsilon$, and set $\widehat\s=\s^{(t+1)}$.

\textbf{Step III (Centered log-scores).}
Let $\widehat s_i$ denote the $i$th entry of $\widehat\s$ and define
\[
\widehat\theta_i
=
\log \widehat s_i
-
\frac{1}{n}\sum_{k=1}^n \log \widehat s_k,
\qquad i=1,\ldots,n.
\]

\textbf{Output:}
$\widehat\s=(\widehat s_1,\ldots,\widehat s_n)^\top$ and
$\widehat\btheta=(\widehat\theta_1,\ldots,\widehat\theta_n)^\top$.
\end{algorithm}

\subsection{Construction of empirical pairwise preferences}
\label{subsec:pref-est}

Both estimators require empirical approximations of the population preference probabilities $p_F(x_i,x_j)$.

Our primary construction is the leave-two-out estimator
\begin{equation}
\label{eq:Rij-def}
\hat p_{ij}^{\mathrm{lto}}
=
\frac{1}{n-2}\sum_{\ell\neq i,j}
\mathbb I\{\delta(x_\ell,x_i)>\delta(x_\ell,x_j)\},
\qquad i\neq j.
\end{equation}
Under the no-ties assumption and the i.i.d.\ sampling scheme,
\[
\mathbb E\!\left[\hat p_{ij}^{\mathrm{lto}} \mid x_i,x_j \right]
=
p_F(x_i,x_j),
\]
so the estimator is conditionally unbiased. It uses all remaining observations as reference points and is statistically efficient in finite samples. 
However, because different pairs share common reference observations, the entries of the empirical preference matrix $\widehat{\mathbf P}$ are dependent.

We also consider a data-reference estimator based on an independent sample $Z_1,\ldots,Z_m\overset{\mathrm{i.i.d.}}{\sim}F$:
\begin{equation}
\label{eq:Rij-ref}
\hat p_{ij}^{\mathrm{ref}}
=
\frac{1}{m}\sum_{l=1}^{m}
\mathbb I\{\delta(Z_l,x_i)>\delta(Z_l,x_j)\}.
\end{equation}
Conditionally on $x_i$ and $x_j$, the indicators 
$\mathbb I\{\delta(Z_l,x_i)>\delta(Z_l,x_j)\}$, $l=1,\ldots,m$, 
are i.i.d.\ Bernoulli random variables with success probability 
$p_F(x_i,x_j)=\mathbb P_{Z\sim F}\{\delta(Z,x_i)>\delta(Z,x_j)\}$.
Consequently,
\[
\mathbb E\!\left[\hat p_{ij}^{\mathrm{ref}}\mid x_i,x_j\right]=p_F(x_i,x_j),
\qquad
\mathrm{Var}\!\left(\hat p_{ij}^{\mathrm{ref}}\mid x_i,x_j\right)
=\frac{p_F(x_i,x_j)\{1-p_F(x_i,x_j)\}}{m}.
\]
Different pairs remain dependent because they share the same reference sample $\{Z_l\}_{l=1}^m$, but this construction separates the reference sample from the comparison sample and avoids the leave-two-out dependence induced by reusing $\{x_\ell\}$ as reference points.

Alternatively, sample splitting may be used to form disjoint comparison and reference sets from the original data. 
While this reduces the effective sample size, it can be convenient in practice.

\begin{remark}[Optional out-of-sample extension]
\label{remark:kernel}
The CORE scores are estimated on the observed sample points $\{x_i\}_{i=1}^n$, which is sufficient for ranking the sample.
If one wishes to evaluate a score at a new point $x\in\mathcal X$, a simple extension is to smooth the fitted scores on the dissimilarity space via kernel weighting.
Given a kernel $K:[0,\infty)\to[0,\infty)$ and bandwidth $h>0$, define
\begin{equation}
\label{eq:kernel-weights}
w_i(x)
=
\frac{K(\delta(x,x_i)/h)}{\sum_{k=1}^n K(\delta(x,x_k)/h)},
\qquad i=1,\ldots,n,
\end{equation}
and set
\begin{equation}
\label{eq:kernel-theta}
\widehat\theta(x)
=
\sum_{i=1}^n w_i(x)\,\widehat\theta_i,
\qquad
\widehat s(x)=\exp\{\widehat\theta(x)\}.
\end{equation}
This yields a Nadaraya--Watson-type extension on $(\mathcal X,\delta)$.
Common choices include a Gaussian kernel $K(t)=\exp(-t^2)$ with a bandwidth selected from the empirical distribution of pairwise distances (e.g., the median), though other choices are possible.
\end{remark}

\begin{remark}[Discrete scores versus parametric scoring]
\label{rem:parametric}
At the population level, $\theta^\star(\cdot)$ is defined on $\mathcal X$ via the M-projection \eqref{eq:theta-star-def}. 
At the sample level, we estimate only its evaluations at the observed data points, yielding $\widehat\btheta\in\mathbb R^n$. 
This discretized strategy avoids imposing additional structural assumptions on $\theta(\cdot)$ and parallels classical depth and rank constructions, which define population functionals but operate on finite samples.

An alternative approach is to estimate $\theta(\cdot)$ directly within a parametric or reproducing-kernel class from pairwise comparisons \citep{FUSE2025}. 
Such methods offer amortized out-of-sample scoring but require stronger modeling assumptions. 
We view these approaches as complementary to the nonparametric M-projection framework developed here.
\end{remark}

\begin{remark}[Win-rate scores and their relation to CORE]
\label{rem:rhat-borda}
A direct sample analogue of \eqref{eq:r-def} is the average win rate
\[
\hat r_j := \frac{1}{n-1}\sum_{i\neq j}\hat p_{ij},
\qquad j=1,\dots,n,
\]
which represents the empirical probability that $x_j$ defeats a randomly chosen opponent. 
At the population level, Theorem~\ref{thm:H-properties} shows that the M-projection score $\theta^\star(\cdot)$ is a strictly monotone transformation of $r_F(\cdot)$. 
Consequently, $\hat r_j$ and $\widehat\theta_j$ induce asymptotically aligned orderings.

The M-projection, however, yields a globally coherent score representation that regularizes noisy pairwise comparisons and provides interpretable log-odds differences, features not available from raw win rates.
\end{remark}

\section{Theoretical Properties}
\label{sec:theory}

This section establishes population and sample-level guarantees for the calibrated centrality
score $\theta^\star$ defined in \eqref{eq:theta-star-def}. Our results clarify (i) when the
population M-projection is well-posed and uniquely defined, (ii) how the resulting score relates
to the intrinsic preference centrality $r_F(\cdot)$ without requiring correct specification of the
BTL working model, and (iii) what structural and statistical properties the calibrated score
inherits, including depth axioms, symmetry-induced orderings, and consistency of the sample
estimator.

\subsection{Existence and uniqueness of the centered minimizer}
\label{subsec:exist-unique}

We begin by establishing that the population M-projection defining $\theta^\star$ is well-defined.
Because the optimization is carried out over a function class, we impose a mild compactness and
convexity condition to guarantee existence and rule out non-identifiability beyond the intrinsic
shift invariance handled by centering.

\begin{assumption}
\label{ass:Theta0}
The centered class $\Theta_0(F)$ is nonempty, convex, and compact in $L^2(F)$.
\end{assumption}

Assumption~\ref{ass:Theta0} is standard in functional M-estimation. Compactness ensures existence of
a minimizer in the infinite-dimensional parameter space, and convexity supports uniqueness once the
objective is shown to be strictly convex on $\Theta_0(F)$.

\begin{theorem}[Existence and uniqueness]
\label{thm:main_result}
Under Assumption~\ref{ass:Theta0}, the risk functional $\mathcal L_F$ defined in
\eqref{eq:population-risk-theta} is strictly convex on $\Theta_0(F)$ and therefore admits a unique minimizer over $\Theta_0(F)$ up to $F$-a.e.\ equality, where $F$-a.e.\ means that the stated property holds outside a measurable set $\mathcal Q \subset\mathcal X$ with $\mathbb P_F(\mathcal Q )=0$.
\end{theorem}

Theorem~\ref{thm:main_result} identifies a single, well-defined calibrated score function
$\theta^\star\in\Theta_0(F)$ as the population target of our methodology.

\begin{remark}[Correct specification as a special case]
\label{rem:identification}
If the working BTL form is correctly specified in the sense that
$p_F(X,Y)=q_{\theta^\star}(X,Y)$ almost surely for some centered score $\theta^\star\in\Theta_0(F)$,
then the M-projection recovers that underlying score function (up to the centering convention).
\end{remark}

\subsection{The monotone calibration link}
\label{subsec:link}

Having established that $\theta^\star$ is uniquely defined, we next show that it provides an
order-preserving reparameterization of the preference centrality $r_F(\cdot)$ introduced in
Section~\ref{sec:intro}. This link is intrinsic to the projection construction and does not rely
on correctness of the BTL working model.

\begin{assumption}
\label{ass:interior}
The calibrated score $\theta^\star$ lies in the relative interior of $\Theta_0(F)$.
\end{assumption}

\begin{theorem}[Monotone calibration link between $\theta^\star$ and $r_F$]
\label{thm:H-properties}
Define
\[
H(t):=\mathbb E_{X\sim F}\big[\sigma\{t-\theta^\star(X)\}\big], \qquad t\in\mathbb R.
\]
Under Assumptions~\ref{ass:Theta0} and \ref{ass:interior}, $H:(-\infty,\infty)\to(0,1)$ is strictly
increasing and satisfies
\begin{equation}
\label{eq:Htheta=r}
H\big(\theta^\star(x)\big)=r_F(x)
\qquad \text{for $F$-a.e.\ }x\in\mathcal X.
\end{equation}
Consequently, $\theta^\star(\cdot)$ is a strictly monotone transformation of $r_F(\cdot)$, and
\[
\theta^\star(x)=H^{-1}\big(r_F(x)\big)
\qquad \text{for $F$-a.e.\ }x\in\mathcal X,
\]
where $H^{-1}$ denotes the inverse of $H$ on its range.
\end{theorem}

Theorem~\ref{thm:H-properties} implies that $\theta^\star$ and $r_F$ induce the same population
ordering up to $F$-null sets: the M-projection preserves the intrinsic preference-based ranking
while providing a coherent one-dimensional score scale.
Assumption~\ref{ass:interior} is a technical condition ensuring that the constrained minimizer is
characterized by first-order optimality conditions (i.e., excluding boundary solutions where only
restricted perturbations are admissible).

The M-projection defines $\theta^\star$ as the unique centered minimizer of a global logistic cross-entropy risk. Equivalently, it minimizes the integrated Kullback–Leibler divergence between the intrinsic preference kernel $p_F(x,y)$ and its BTL-type approximation $q_\theta(x,y)=\sigma\{\theta(x)-\theta(y)\}$. Thus, $\theta^\star$ is the closest logit-consistent representation of $p_F$ within the BTL family.

Although $\theta^\star$ and $r_F$ induce identical population orderings, $\theta^\star$ additionally provides a coherent log-odds scale satisfying
\[
q_{\theta^\star}(x,y)
=
\sigma\{\theta^\star(x)-\theta^\star(y)\}.
\]
This probabilistic representation turns preference aggregation into a convex M-projection problem with a unique target, which facilitates estimation and finite-sample analysis.

\subsection{Depth properties of the M-projection scores}
\label{subsec:depth}

The previous subsection established that the calibrated score
$\theta^\star$ is uniquely defined and is a strictly monotone
reparameterization of the intrinsic preference centrality $r_F(\cdot)$.
We now investigate whether the logistic M-projection preserves the
center--outward geometry encoded in the underlying preference relation.

Statistical depth provides an axiomatic framework for formalizing
center--outward structure.  We adopt the classical affine depth axioms
of \citet{zuo2000general}.  Since these axioms are formulated through
affine transformations and linear interpolation, we restrict in this
subsection to the Euclidean setting $\mathcal X=\mathbb R^d$ and treat
$F$ as a distribution on $\mathbb R^d$.
The dissimilarity $\delta$ that defines the preference kernel
$p_F(x,y)$ is allowed to be any measurable dissimilarity on
$\mathbb R^d$; it need not coincide with the Euclidean norm.

Rather than assuming that $r_F(\cdot)$ is itself a depth function, we
impose structural regularity directly on the preference geometry.
The following conditions encode center--outward behavior at the level
of the preference kernel.

\begin{assumption}[Center--outward regularity of the preference geometry]
\label{ass:depth}
The preference kernel $p_F(x,y)$ satisfies:

\begin{enumerate}
\item[(a)] \textit{Affine equivariance.}
For every invertible affine map $T_{A,b}(x)=Ax+b$ and
$G=T_{A,b}\#F$,
\[
p_G(T_{A,b}x,T_{A,b}y)=p_F(x,y)
\quad \text{for all } x,y\in\mathbb R^d ,
\]
and 
\[
\Theta_0(G)
=
\{\theta\circ T_{A,b}^{-1}:\theta\in\Theta_0(F)\}.
\]

\item[(b)] \textit{Existence of a preference center and center--outward ordering.}
There exists $\mu\in\mathbb R^d$ such that
\[
r_F(\mu)=\sup_{x\in\mathbb R^d} r_F(x),
\]
and for all $x\in\mathbb R^d$ and $\alpha\in[0,1]$,
\[
r_F((1-\alpha)\mu+\alpha x)\ge r_F(x).
\]

\item[(c)] \textit{Vanishing preference at $\delta$-infinity.}
If $\delta(y_m,\mu)\to\infty$, then
\[
r_F(y_m)\to 0 .
\]
\end{enumerate}
\end{assumption}

Assumption~\ref{ass:depth} requires that the underlying preference
relation exhibits affine invariance and center--outward regularity.

\begin{assumption}[Euclidean regularity and continuity of $\delta$]
\label{ass:euclidean-reg}
Assume that $\Theta_0(F) \subset L_0^2(F) \cap C(\R^d)$, where $C(\mathbb R^d)$ denotes the space of continuous real-valued functions on $\mathbb R^d$.
The distribution $F$ is absolutely continuous with respect to the Lebesgue measure and has full support on $\mathbb R^d$.
Moreover, for $F$-a.e.\ $z\in\mathbb R^d$, the map $y\mapsto \delta(z,y)$ is continuous on $\mathbb R^d$.
\end{assumption}

Assumption~\ref{ass:euclidean-reg} ensures that the M-projection can be interpreted pointwise. 
In the $L^2(F)$ formulation, two scores that differ only on an $F$-null set are identified, so the value of the score at a specific location is not determined. 
This is problematic because several depth axioms evaluate the score at fixed spatial points, such as the center in (D2). 
Restricting $\Theta_0(F)$ to continuous functions removes this ambiguity and turns $F$-a.e.\ calibration identities into pointwise statements.

The continuity of $\delta$ implies that $p_F(x,y)$ and $r_F(y)$ vary continuously with their arguments. 
If $\delta$ were discontinuous, the induced preference structure could exhibit jump discontinuities that are incompatible with a continuous score $\theta\in\Theta_0(F)$. 
In that case, the minimizer may lie on the boundary of the parameter space, so the interior first-order calibration condition would no longer apply.

Absolute continuity and full support of $F$ ensure that the calibration relation identifies the continuous score on all of $\mathbb R^d$. 
Without full support, the score is not determined outside the support of $F$, and global geometric properties such as vanishing at infinity cannot be stated meaningfully.

\begin{theorem}[Geometry preservation under logistic M-projection]
\label{th:mproj-depth}
Suppose Assumptions~\ref{ass:Theta0},
\ref{ass:interior}, \ref{ass:depth} and \ref{ass:euclidean-reg} hold.
Then, for all $x \in \mathbb R^d$, the calibrated score $\theta^\star$ satisfies:

\begin{enumerate}
\item[(D1)] \textbf{Affine equivariance (up to centering).}
For $G=T_{A,b}\#F$,
\[
\theta_G^\star(T_{A,b}x)
=
\theta_F^\star(x).
\]

\item[(D2)] \textbf{Maximality at the center.}
If $\mu$ is as in Assumption~\ref{ass:depth}(b), then
\[
\theta^\star(\mu)\ge \theta^\star(x).
\]

\item[(D3)] \textbf{Monotonicity toward the center.}
For $x_\alpha=(1-\alpha)\mu+\alpha x$ with $\alpha \in [0,1]$,
\[
\theta^\star(x_\alpha)\ge \theta^\star(x).
\]

\item[(D4)] \textbf{Behavior at $\delta$-infinity.}
If $\delta(y_m,\mu)\to\infty$, then
\[
\theta^\star(y_m)\to -\infty .
\]
\end{enumerate}
\end{theorem}

Theorem~\ref{th:mproj-depth} shows that the logistic M-projection
preserves the intrinsic center--outward ordering encoded in the
preference geometry.  Although $\theta^\star$ is defined as the
minimizer of a parametric cross-entropy risk, it retains the full
affine depth structure whenever the underlying preference relation
exhibits such geometry.

Since $s^\star(x)=\exp\{\theta^\star(x)\}$ is a strictly monotone
transformation of $\theta^\star(x)$, the same ordering and
center--outward properties hold for the strength score.

\subsubsection{One-dimensional symmetric distribution}
\label{subsec:1d-radial-r}

We now show that the regularity conditions in Assumption~\ref{ass:depth}
are naturally satisfied in classical symmetric benchmark models.
This clarifies that the center--outward structure assumed above
is not restrictive, but instead reflects the geometry
of familiar distributions.

Consider the one-dimensional Euclidean setting $\mathcal X=\mathbb R$.
Under symmetry about a location parameter $\mu$,
any reasonable population centrality must assign equal importance to
$\mu+t$ and $\mu-t$ and depend on $x$
only through its distance from the center.
This intuition underlies classical univariate depth constructions.

\begin{assumption}[Reflection symmetry]
\label{ass:symm-1d-r}
Assume $\mathcal X=\mathbb R$ and $\delta(x,y)=|x-y|$.
Suppose that $F$ is symmetric about $\mu$ such that $X\sim F$,
\[
X-\mu \overset{d}{=} \mu-X.
\]
\end{assumption}

Under this symmetry, the induced preference centrality $r_F(\cdot)$
is automatically maximized at $\mu$,
is symmetric around $\mu$,
and decreases as $|x-\mu|$ increases.
Consequently, Assumption~\ref{ass:depth}(b)--(c)
are satisfied.

\begin{proposition}
\label{prop:1d-radial-theta}
Suppose Assumptions~\ref{ass:Theta0}, \ref{ass:interior}
and~\ref{ass:symm-1d-r} hold.
Then there exists a function
$\varphi:[0,\infty)\to\mathbb R$ such that
\[
\theta^\star(x)=\varphi(|x-\mu|),\qquad x\in\mathbb R.
\]
If in addition Assumption~\ref{ass:depth}(b) holds,
$\varphi$ can be chosen non-increasing.
In particular,
if $|x-\mu|<|y-\mu|$, then
\[
\theta^\star(x)\ge \theta^\star(y).
\]
\end{proposition}

Proposition~\ref{prop:1d-radial-theta} shows that,
under reflection symmetry,
the calibrated population score depends only on distance to the center
and decreases monotonically as one moves away from $\mu$.
Thus, the M-projection preserves the intrinsic symmetry
and recovers the canonical one-dimensional center--outward ordering.

\subsubsection{Multivariate elliptically symmetric distribution}
\label{subsec:multivariate-ordering}

We next consider the multivariate Euclidean setting
$\mathcal X=\mathbb R^d$ under elliptical symmetry.
This setting generalizes reflection symmetry
to higher dimensions and encompasses
Gaussian and more general elliptical distributions.

Under elliptical symmetry,
centrality is naturally characterized
through distances measured relative to
a positive definite scatter matrix $\Sigma$.
Classical multivariate depth constructions,
including Mahalanobis depth,
induce center--outward orderings
based on ellipsoidal level sets.

\begin{assumption}
\label{ass:symm-Rd-maha}
Assume $\mathcal X=\mathbb R^d$ and equip it with the Mahalanobis dissimilarity
\[
\delta_\Sigma(x,y)
=
\bigl\{(x-y)^\top\Sigma^{-1}(x-y)\bigr\}^{1/2},
\qquad \Sigma \text{ positive definite}.
\]
Suppose that for $X\sim F$ and every orthogonal matrix $O$,
\[
\Sigma^{-1/2}(X-\mu)
\overset{d}{=}
O\,\Sigma^{-1/2}(X-\mu),
\]
so that $F$ is elliptically symmetric about $\mu$
with scatter matrix $\Sigma$.
\end{assumption}

Under this condition,
the preference centrality $r_F(\cdot)$
is constant on ellipsoids centered at $\mu$
and decreases as the Mahalanobis radius
$\delta_\Sigma(x,\mu)$ increases.
Therefore Assumption~\ref{ass:depth}(b)--(c)
are again satisfied.

\begin{proposition}
\label{prop:Rd-radial-theta}
Suppose Assumptions~\ref{ass:Theta0}, \ref{ass:interior}
and~\ref{ass:symm-Rd-maha} hold.
Then there exists a function
$\varphi:[0,\infty)\to\mathbb R$ such that
\[
\theta^\star(x)
=
\varphi\bigl(\delta_\Sigma(x,\mu)\bigr),
\qquad x\in\mathbb R^d.
\]
If in addition Assumption~\ref{ass:depth}(b) holds,
$\varphi$ can be chosen non-increasing.
In particular,
if $\delta_\Sigma(x,\mu)<\delta_\Sigma(y,\mu)$,
then
\[
\theta^\star(x)\ge \theta^\star(y).
\]
\end{proposition}

Proposition~\ref{prop:Rd-radial-theta} shows that,
under elliptical symmetry,
the calibrated score induces
the same center--outward ordering
as Mahalanobis depth.
Hence the logistic M-projection
preserves the intrinsic geometric structure
encoded by the covariance matrix $\Sigma$,
rather than distorting it.

\subsection{Consistency of the sample M-estimator}
\label{subsec:consistency}

We now turn to statistical properties of the sample-level estimator introduced in Section~\ref{subsec:core_gd}.  Our goal is to show that the empirical minimizer of the cross-entropy risk based on estimated pairwise preferences consistently recovers the population calibrated score $\theta^\star$ evaluated at the observed data points.

Importantly, the BTL form is used purely as a working projection family.  No correct specification assumption is imposed; the target remains the population M-projection $\theta^\star$ defined in \eqref{eq:theta-star-def}.

\begin{assumption}[Essential boundedness of the population score]
\label{ass:bounded}
There exists $B_\star<\infty$ such that
\[
\|\theta^\star\|_{L^\infty(F)} \le B_\star .
\]
\end{assumption}

Assumption~\ref{ass:bounded} requires boundedness only in the $F$-essential sense.  
It does not preclude $\theta^\star(x)\to -\infty$ as $\delta(x,\mu)\to\infty$ outside regions of non-negligible probability mass.  
Rather, it ensures that the population target evaluated at random draws $X\sim F$ lies in a bounded range almost surely.

This assumption is imposed solely to facilitate finite-sample analysis.  
It allows us to localize the empirical optimization problem to a compact parameter set, under which the logistic risk becomes uniformly strongly convex and the empirical process can be controlled via standard concentration arguments.  
Such localization is common in nonparametric M-estimation and does not alter the population target.

We consider two plug-in constructions of the empirical preference matrix:
the leave-two-out estimator $\widehat{\mathbf P}^{\mathrm{lto}}$ in
\eqref{eq:Rij-def}, and the reference-sample estimator
$\widehat{\mathbf P}^{\mathrm{ref}}$ in \eqref{eq:Rij-ref} based on an
independent reference sample of size $m$.

For theoretical analysis, we restrict the optimization to the bounded centered set
\[
\Theta_n(B)
=
\Big\{\btheta\in\mathbb R^n:\ \mathbf 1^\top\btheta=0,\ \|\btheta\|_\infty\le B\Big\},
\]
where $B>3B_\star$ is fixed.  We define
\[
\hat\btheta^{\mathrm{lto}}
\in
\argmin_{\btheta\in\Theta_n(B)}
\mathcal L_n(\btheta;\widehat{\mathbf P}^{\mathrm{lto}}),
\qquad
\hat\btheta^{\mathrm{ref}}
\in
\argmin_{\btheta\in\Theta_n(B)}
\mathcal L_n(\btheta;\widehat{\mathbf P}^{\mathrm{ref}}).
\]

\begin{theorem}[Empirical $L^2$ consistency]
\label{thm:plugin-consistency}
Suppose Assumptions~\ref{ass:Theta0}, \ref{ass:interior}, and \ref{ass:bounded} hold. Then each optimization problem above admits a unique minimizer over $\Theta_n(B)$.
Moreover,
\[
\frac{1}{n}\sum_{i=1}^n
\big(\hat\theta_i^{\mathrm{lto}} - \theta^\star(X_i)\big)^2
=
O_{\mathbb P}\!\left(\frac{\log n}{n}\right),
\]
and for an independent reference sample of size $m$,
\[
\frac{1}{n}\sum_{i=1}^n
\big(\hat\theta_i^{\mathrm{ref}} - \theta^\star(X_i)\big)^2
=
O_{\mathbb P}\!\left(\frac{1}{n} + \frac{\log n}{m}\right).
\]
Here $A_n = O_{\mathbb P}(a_n)$ means that $A_n/a_n$ is bounded in probability.
\end{theorem}

Theorem~\ref{thm:plugin-consistency} establishes that the empirical projection consistently recovers the population calibrated score evaluated at the observed sample points.
The resulting error can be decomposed into two components:
the statistical error from estimating the pairwise preference matrix,
and the intrinsic $n^{-1}$ term arising from the strong convexity of the empirical logistic objective.
More generally, for any empirical preference matrix
$\widehat{\mathbf P}=\{\hat p_{ij}\}$ satisfying
$\hat p_{ij}\in[0,1]$ and $\hat p_{ij}+\hat p_{ji}=1$,
the empirical minimizer over $\Theta_n(B)$ exists uniquely and satisfies
\[
\frac{1}{n}\sum_{i=1}^n
(\hat\theta_i-\theta^\star(X_i))^2
=
O_{\mathbb P}\!\Big(
n^{-1}
+
\|\widehat{\mathbf P}-\mathbf P\|_{\max}^2
\Big),
\]
where
\[
\|\widehat{\mathbf P}-\mathbf P\|_{\max}
=
\max_{i\ne j}|\,\hat p_{ij}-p_{ij}\,|.
\]
This bound shows that the projection step itself does not introduce additional statistical complexity: the overall error is governed by the accuracy of preference estimation together with the intrinsic parametric $n^{-1}$ term.
In particular, when the preference matrix is estimated sufficiently accurately, the $n^{-1}$ term dominates, yielding the optimal parametric rate for finite-dimensional M-estimation.

A further notable feature is that the rate is dimension-free, depending only on sample size and preference-estimation error rather than on the ambient dimension or geometric complexity of the observation space.
This highlights a key advantage of the proposed framework in high-dimensional or non-Euclidean settings, where many geometric or depth-based procedures suffer from dimensional degradation.

\begin{remark}[Spectral approximation]
Section~\ref{subsec:core_gd} also introduced a spectral estimator based on the stationary distribution of a Markov operator constructed from $\widehat{\mathbf P}$. 
This estimator can be viewed as a linearized approximation to the logistic M-projection studied above.
A separate finite-sample analysis of the spectral estimator is beyond the scope of the present paper.
\end{remark}


\section{Simulation Studies}
\label{sec:sim}

This section investigates the finite-sample behavior of the proposed CORE estimators under Euclidean geometry. The simulations are designed to validate the structural and statistical properties
established in Section~\ref{sec:theory}. In particular, we examine:
(i) whether the empirical estimators recover the population M-projection $\theta^\star$,
(ii) whether the induced ranking aligns with the intrinsic preference centrality $r_F(\cdot)$,
and (iii) how CORE compares with classical depth-based procedures across
different geometric regimes.

\subsection{One-dimensional score convergence}

We begin with a one-dimensional setting in which the population preference function
$p_F(\cdot,\cdot)$ admits a closed-form expression.
This allows us to directly visualize the estimated score functions and assess their
convergence toward the population target as the sample size increases.

We consider sample sizes $n\in\{50,200,500,2000\}$ with $20$ independent replications
for each $n$.
In each replication, we generate $x_1,\ldots,x_n \overset{\mathrm{i.i.d.}}{\sim} F$, where
$F$ is either (i) the standard normal distribution $N(0,1)$, or
(ii) a skewed Laplace distribution with density
\[
f(x)=\frac{1}{b_L+b_R}\{\exp(x/b_L)\mathbf 1(x<0)+\exp(-x/b_R)\mathbf 1(x\ge 0)\},
\]
with $(b_L,b_R)=(2,1)$.
The dissimilarity is set to $\delta(x,y)=|x-y|$.

We compare three constructions of the pairwise preference matrix.
First, we consider the \emph{population} matrix $\mathbf P^{\mathrm{pop}}=\{p_{ij}\}_{i,j=1}^n$ with
$p_{ij}=p_F(x_i,x_j)$.
In one dimension, letting $F_{\mathrm{cdf}}$ denote the distribution function of $F$,
the preference probability admits the closed form
\[
p_F(x_i,x_j)=
\begin{cases}
1 - F_{\mathrm{cdf}}\big((x_i+x_j)/2\big), & x_i<x_j,\\[0.6em]
F_{\mathrm{cdf}}\big((x_i+x_j)/2\big), & x_i>x_j.
\end{cases}
\]

Second, we use the independent \emph{reference-sample} estimator
$\widehat{\mathbf P}^{\mathrm{ref}}=\{\hat p_{ij}^{\mathrm{ref}}\}_{i,j=1}^n$ defined in \eqref{eq:Rij-ref},
computed from an independent sample $Z_1,\ldots,Z_n\overset{\mathrm{i.i.d.}}{\sim}F$.
Third, we consider the \emph{leave-two-out} estimator
$\widehat{\mathbf P}^{\mathrm{lto}}=\{\hat p_{ij}^{\mathrm{lto}}\}_{i,j=1}^n$ in \eqref{eq:Rij-def},
which uses the observed sample itself as reference points.

For each preference matrix, we compute the corresponding score vector using the
gradient-based implementation (Algorithm~\ref{al:GD-CORE}),
referred to as \textit{Population-GD}, \textit{Reference-GD}, and \textit{Leave-out-GD},
respectively.
In addition, we apply the spectral estimator (Algorithm~\ref{al:CORE-spectral}) to
$\widehat{\mathbf P}^{\mathrm{lto}}$, yielding the \textit{Leave-out-Spectral} estimator.

\begin{figure}[htpb!]
    \centering    \includegraphics[width=0.99\linewidth]{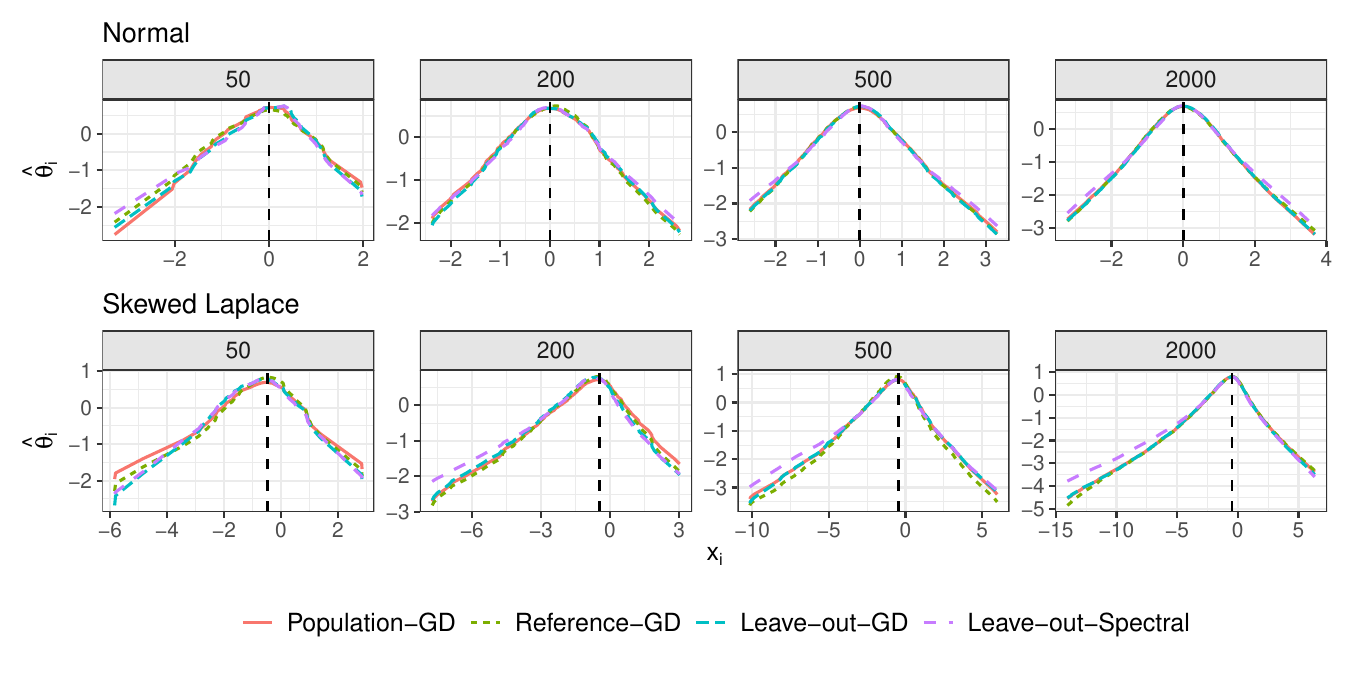}
\caption{
One-dimensional comparison of four CORE estimators.
Each panel corresponds to a distribution (Normal or skewed Laplace) and a sample size
$n\in\{50,200,500,2000\}$.
Estimated scores $\hat\theta_i$ are plotted against the sample points $x_i$.
The vertical dashed line marks the population maximizer $\mu^\star$ of $r_F(\cdot)$.
}
    \label{fig:theta_1D_method}
\end{figure}

Figure~\ref{fig:theta_1D_method} overlays the four estimated score curves obtained from
the same dataset for each combination of distribution and sample size.
As $n$ increases, the three gradient-based estimators become nearly indistinguishable,
indicating that the leave-two-out and reference constructions recover the same limiting
score as the population benchmark.
The spectral estimator exhibits slightly larger deviations in the tails, but induces
essentially the same ordering.

\begin{figure}[t!]
    \centering
    \includegraphics[width=0.95\linewidth]{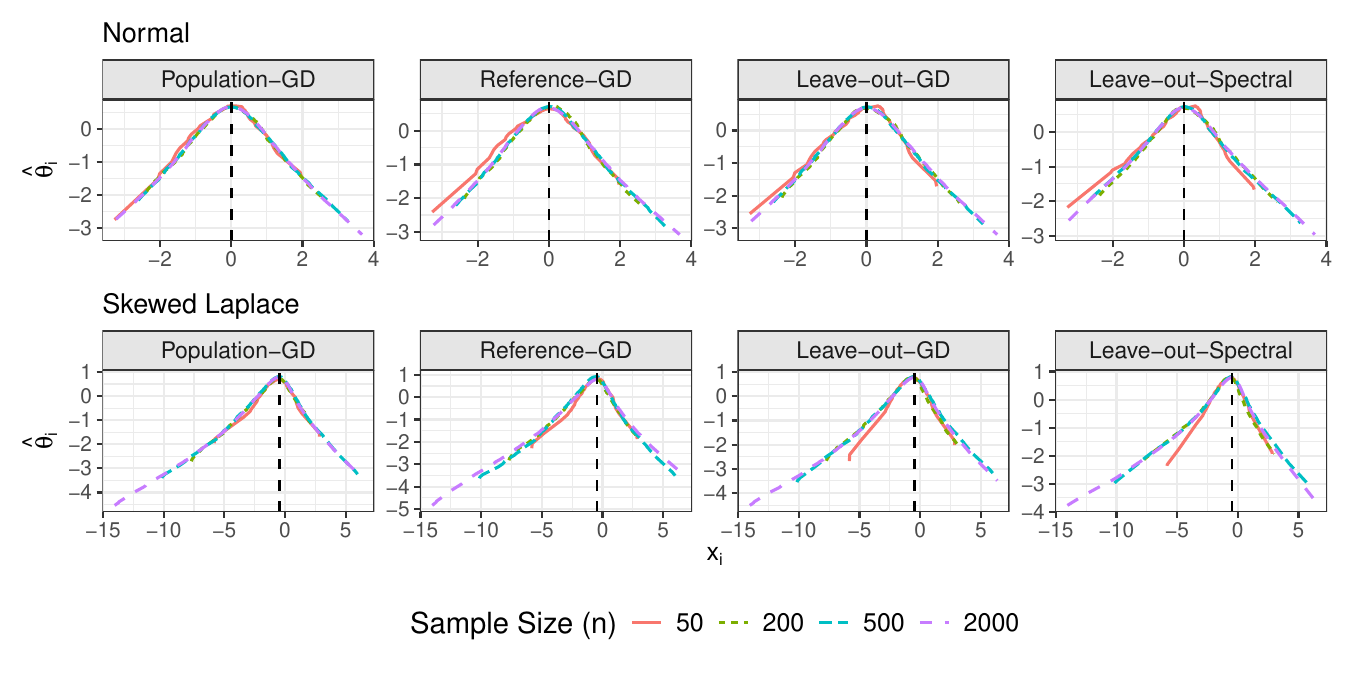}
\caption{
Comparison of the four estimators.
Each panel fixes a distribution (Normal or skewed Laplace), and overlays the estimated score curves obtained at sample sizes $n\in\{50,200,500,2000\}$.
In each panel, the scores $\hat\theta_i$ are plotted against the ordered sample points $x_{(i)}$.
The vertical dashed line marks the population maximizer $\mu^\star=0$ for the Normal case and $\mu^\star=-0.462$ for the skewed Laplace case.
}
    \label{fig:theta_1D_convergence}
\end{figure}

\begin{figure}[t!]
\centering
\includegraphics[width=0.95\linewidth]{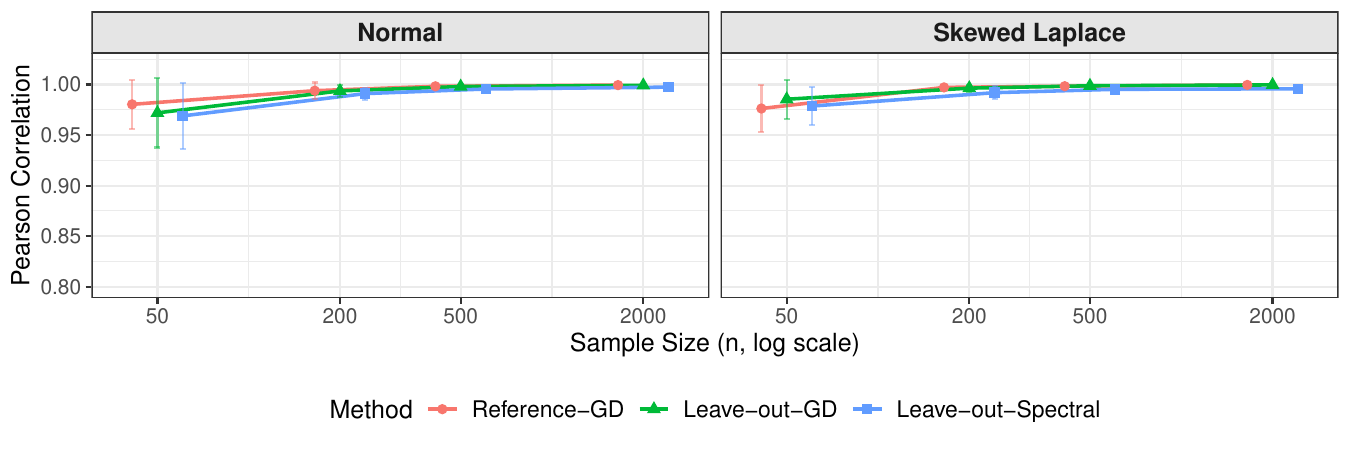}
\caption{
Mean Pearson correlation between the estimated score vectors and the Population-GD benchmark
over $20$ independent replications.
Points indicate the mean correlation and vertical error bars denote $\pm 1$ standard deviation.
}
\label{fig:mean_correlation}
\end{figure}

Figure~\ref{fig:theta_1D_convergence} fixes the distribution and estimation method and
overlays the score curves across different sample sizes.
For all methods, the curves stabilize toward a common limiting shape as $n$ grows.
The vertical dashed line marks the population maximizer $\mu^\star$ of $r_F(\cdot)$
($\mu^\star=0$ for the normal distribution and $\mu^\star=-0.462$ for the skewed
Laplace case).
Even for moderate sample sizes, the estimated scores peak near $\mu^\star$,
demonstrating reliable recovery of the population center.

Figure~\ref{fig:mean_correlation} reports the mean Pearson correlation between each
estimated score vector and the Population-GD benchmark across $R=20$ replications.
As $n$ increases, correlations for all methods approach one and the variability
across replications diminishes.
Notably, the gap between Reference-GD and Leave-out-GD narrows rapidly with $n$,
indicating that the leave-two-out estimator achieves comparable accuracy without
requiring an additional independent reference sample.
The spectral estimator remains slightly less correlated than the gradient-based
methods, but consistently exhibits high agreement with the population benchmark.

\subsection{Rank recovery in moderate and high dimensions}
\label{subsec:sim_hd_depth}

We next evaluate the ability of CORE and competing procedures to recover the
population ranking induced by the preference-based centrality functional
$r_F(\cdot)$ under Euclidean dissimilarity $\delta(x,y)=\|x-y\|_2$ in moderate and high dimensions.

We consider four $(n,d)$ settings:
$(150,80)$ and $(500,200)$ for the $n>d$ regime, and
$(80,200)$ and $(150,500)$ for the high-dimensional regime $n<d$.
All results are averaged over $20$ independent replicates.

Data are generated from three representative models:
(i) a multivariate Student's $t_\nu$ distribution with $\nu=5$ and identity scatter;
(ii) a balanced Gaussian mixture
$0.5,N(u,I_d)+0.5,N(-u,I_d)$, where
$u=(4,\ldots,4,0,\ldots,0)^\top$ has value $4$ on the first $10$ coordinates;
and (iii) a coordinate-wise skewed Laplace model
$\mathrm{ALD}(0,2,10)$, with the first coordinate scaled by $100$
to induce strong skewness and anisotropy.

For each replicate, we construct the leave-two-out pairwise probability matrix
$\widehat{\mathbf P}^{\mathrm{lto}}$ defined in \eqref{eq:Rij-def}.
CORE-GD reports the centered score vector $\widehat\btheta$,
whereas CORE-Spectral reports the stationary score vector $\widehat{\s}$.




When $n>d$, we compare against four classical depth-based scores:
halfspace depth, Mahalanobis depth, projection depth, and spatial depth,
all implemented using the \texttt{ddalpha} R package \citep{JSSv091i05}.
When $n<d$, many classical depth notions become degenerate or computationally infeasible.
We therefore include two high-dimensional alternatives:
randomized-projection spatial depth (RP-Spatial),
which approximates spatial depth via aggregation over random low-dimensional projections
\citep{cuesta2008random},
and the negative $\ell_2$ distance to the sample mean (Neg-L2).

Performance is evaluated using the Spearman rank correlation between each method’s
estimated score vector and the benchmark ranking
$(r_F(X_1),\ldots,r_F(X_n))$.
Since $r_F(\cdot)$ has no closed-form expression in general, we approximate
$r_F(X_i)$ by Monte Carlo.
Specifically, for each $y=X_i$, we draw
$X'_1,\ldots,X'_{M_1}\overset{\mathrm{i.i.d.}}{\sim}F$
and $Z_1,\ldots,Z_{M_2}\overset{\mathrm{i.i.d.}}{\sim}F$, and compute
\[
\hat r_F(y)
=
\frac{1}{M_1M_2}\sum_{j=1}^{M_2}\sum_{k=1}^{M_1}
\mathbb{I}\left(\|Z_j-X'_k\|_2>\|Z_j-y\|_2\right).
\]
We take $M_1=2000$ and $M_2=10000$ throughout.

\begin{table}[ht]
\centering
\caption{Spearman correlation ($\text{Mean}_{\text{SD}}$ over $20$ replicates) between each method's score and Monte Carlo $\hat r_F(X_i)$ for $(n,d) \in \{(150, 80), (500, 200)\}$.}
\label{tab:dn_lt_n}
\begin{tabular}{lcccccc}
\toprule
& \multicolumn{2}{c}{t} & \multicolumn{2}{c}{Gaussian Mixture} & \multicolumn{2}{c}{Skewed Laplace} \\
\cmidrule(lr){2-3}\cmidrule(lr){4-5}\cmidrule(lr){6-7}
Method
& $(150,80)$ & $(500,200)$
& $(150,80)$ & $(500,200)$
& $(150,80)$ & $(500,200)$ \\
\midrule
CORE-GD         & $0.999_{0.000}$ & $1.000_{0.000}$ & $0.962_{0.029}$ & $0.989_{0.009}$ & $0.980_{0.013}$ & $0.992_{0.007}$ \\
CORE-Spectral   & $0.996_{0.002}$ & $0.996_{0.001}$ & $0.971_{0.020}$ & $0.990_{0.007}$ & $0.981_{0.013}$ & $0.992_{0.007}$ \\
Halfspace       & $0.800_{0.036}$ & $0.930_{0.009}$ & $0.143_{0.009}$ & $0.222_{0.037}$ & $0.164_{0.121}$ & $0.181_{0.075}$ \\
Mahalanobis     & $0.959_{0.010}$ & $0.989_{0.001}$ & $0.401_{0.066}$ & $0.556_{0.024}$ & $0.073_{0.099}$ & $0.101_{0.049}$ \\
Projection      & $0.945_{0.012}$ & $0.949_{0.007}$ & $0.375_{0.065}$ & $0.306_{0.046}$ & $0.482_{0.068}$ & $0.502_{0.056}$ \\
Spatial         & $0.959_{0.010}$ & $0.989_{0.001}$ & $0.401_{0.067}$ & $0.556_{0.025}$ & $0.073_{0.099}$ & $0.101_{0.049}$ \\
\bottomrule
\end{tabular}
\end{table}


\begin{table}[ht]
\centering
\caption{Spearman correlation ($\text{Mean}_{\text{SD}}$ over $20$ replicates) between each method's score and Monte Carlo $\hat r_F(X_i)$ for the $d > n$ settings ($(n,d) \in \{(80, 200), (150, 500)\}$).}
\label{tab:dn_gt_n}
\begin{tabular}{lcccccc}
\toprule
& \multicolumn{2}{c}{t} & \multicolumn{2}{c}{Gaussian Mixture} & \multicolumn{2}{c}{Skewed Laplace} \\
\cmidrule(lr){2-3}\cmidrule(lr){4-5}\cmidrule(lr){6-7}
Method
& $(80,200)$ & $(150,500)$
& $(80,200)$ & $(150,500)$
& $(80,200)$ & $(150,500)$ \\
\midrule
CORE-GD         & $0.999_{0.000}$ & $1.000_{0.000}$ & $0.953_{0.043}$ & $0.961_{0.040}$ & $0.970_{0.024}$ & $0.984_{0.012}$ \\
CORE-Spectral   & $0.990_{0.007}$ & $0.992_{0.008}$ & $0.962_{0.031}$ & $0.970_{0.030}$ & $0.973_{0.020}$ & $0.983_{0.013}$ \\
Neg-L2          & $0.999_{0.000}$ & $1.000_{0.000}$ & $0.786_{0.176}$ & $0.840_{0.140}$ & $0.795_{0.079}$ & $0.739_{0.081}$ \\
RP-Spatial      & $0.989_{0.003}$ & $0.990_{0.003}$ & $0.691_{0.095}$ & $0.579_{0.088}$ & $0.911_{0.064}$ & $0.856_{0.065}$ \\
\bottomrule
\end{tabular}
\end{table}

Tables~\ref{tab:dn_lt_n} and~\ref{tab:dn_gt_n} report the resulting Spearman correlations
between each method’s score and the Monte Carlo approximation $\hat r_F(X_i)$.
Table~\ref{tab:dn_lt_n} summarizes the $n>d$ settings.
Across all distributions and sample sizes, both CORE-GD and CORE-Spectral exhibit consistently high rank agreement with the benchmark,
with correlations close to one.
The spectral approximation performs comparably to the gradient-based estimator,
indicating that it captures the same underlying preference structure while avoiding
iterative optimization.
In contrast, classical depth-based methods show strong agreement mainly under the
elliptically symmetric Student’s $t$ distribution, with substantially weaker concordance
under the Gaussian mixture and skewed Laplace models.
This behavior is expected, as these depth notions target different population concepts
of centrality than the preference-based functional $r_F(\cdot)$.

Table~\ref{tab:dn_gt_n} presents the results for the $n<d$ regime.
Both CORE estimators continue to exhibit strong agreement with the benchmark
ranking even when the dimension exceeds the sample size.
The Neg-L2 baseline performs well primarily under spherical symmetry, consistent with
the analysis in Section~\ref{subsec:multivariate-ordering}, but becomes less informative
under multimodality and skewness.
RP-Spatial partially mitigates high-dimensional degeneracy through random projections,
yet still shows reduced agreement relative to CORE in non-elliptical settings.
Overall, the results illustrate that CORE maintains stable rank recovery across
a wide range of distributional structures and dimensional regimes.

\begin{table}[t!]
\centering
\caption{Spearman correlation ($\text{Mean}_{\text{SD}}$ over $20$ replicates) between each method's score and the true log-density $\log f(X_i)$ for $(n,d) \in \{(150, 80), (500, 200)\}$.}
\label{tab:dens_dn_lt_n}
\resizebox{\textwidth}{!}{%
\begin{tabular}{lcccccc}
\toprule
& \multicolumn{2}{c}{t} & \multicolumn{2}{c}{Gaussian Mixture} & \multicolumn{2}{c}{Skewed Laplace} \\
\cmidrule(lr){2-3}\cmidrule(lr){4-5}\cmidrule(lr){6-7}
Method & $(150,80)$ & $(500,200)$ & $(150,80)$ & $(500,200)$ & $(150,80)$ & $(500,200)$ \\
\midrule
CORE-GD & $0.999_{0.000}$ & $1.000_{0.000}$ & $0.622_{0.037}$ & $0.698_{0.030}$ & $0.074_{0.079}$ & $0.087_{0.039}$ \\
CORE-Spectral & $0.996_{0.001}$ & $0.996_{0.001}$ & $0.629_{0.036}$ & $0.702_{0.030}$ & $0.072_{0.080}$ & $0.083_{0.039}$ \\
Halfspace & $0.810_{0.032}$ & $0.927_{0.007}$ & $0.141_{0.021}$ & $0.206_{0.044}$ & $0.274_{0.066}$ & $0.238_{0.035}$ \\
Mahalanobis & $0.961_{0.007}$ & $0.989_{0.001}$ & $0.654_{0.049}$ & $0.749_{0.026}$ & $0.619_{0.049}$ & $0.636_{0.025}$ \\
Projection & $0.943_{0.014}$ & $0.947_{0.005}$ & $0.475_{0.062}$ & $0.360_{0.031}$ & $0.377_{0.073}$ & $0.306_{0.037}$ \\
Spatial & $0.961_{0.008}$ & $0.989_{0.001}$ & $0.654_{0.049}$ & $0.749_{0.026}$ & $0.618_{0.049}$ & $0.636_{0.025}$ \\
\bottomrule
\end{tabular}%
}
\end{table}

\begin{table}[t!]
\centering
\caption{Spearman correlation ($\text{Mean}_{\text{SD}}$ over $20$ replicates) between each method's score and the true log-density $\log f(X_i)$ for $(n,d) \in \{(80, 200), (150, 500)\}$.}
\label{tab:dens_dn_gt_n}
\begin{tabular}{lcccccc}
\toprule
& \multicolumn{2}{c}{t} & \multicolumn{2}{c}{Gaussian Mixture} & \multicolumn{2}{c}{Skewed Laplace} \\
\cmidrule(lr){2-3}\cmidrule(lr){4-5}\cmidrule(lr){6-7}
Method
& $(80,200)$ & $(150,500)$
& $(80,200)$ & $(150,500)$
& $(80,200)$ & $(150,500)$ \\
\midrule
CORE-GD       & $0.999_{0.000}$ & $1.000_{0.000}$ & $0.666_{0.057}$ & $0.770_{0.045}$ & $0.119_{0.096}$ & $0.110_{0.063}$ \\
CORE-Spectral & $0.990_{0.007}$ & $0.992_{0.008}$ & $0.679_{0.062}$ & $0.781_{0.040}$ & $0.117_{0.094}$ & $0.107_{0.064}$ \\
Neg-L2        & $0.999_{0.000}$ & $1.000_{0.000}$ & $0.468_{0.108}$ & $0.636_{0.110}$ & $0.067_{0.106}$ & $0.043_{0.057}$ \\
RP-Spatial    & $0.989_{0.003}$ & $0.990_{0.003}$ & $0.481_{0.085}$ & $0.463_{0.067}$ & $0.101_{0.109}$ & $0.066_{0.059}$ \\
\bottomrule
\end{tabular}
\end{table}

Tables~\ref{tab:dens_dn_lt_n} and~\ref{tab:dens_dn_gt_n} report Spearman correlations between
the estimated scores and the true log-density values $\log f(X_i)$.
Under the symmetric Student’s $t$ model, CORE is almost perfectly correlated with
$\log f(X_i)$ in both dimensional regimes, consistent with
Proposition~\ref{prop:Rd-radial-theta}, which shows that the population score reduces to a
radial function of the Mahalanobis distance in the elliptical case.
In non-elliptical settings, correlations with $\log f$ can be substantially lower,
reflecting the fact that different procedures target distinct notions of centrality.
In particular, CORE is driven by the global geometry induced by the Euclidean
dissimilarity, whereas $\log f$ represents a local quantity that is sensitive to
multimodality, skewness, and anisotropic scaling.

We also compare computation time across methods.
CORE-Spectral is consistently faster than CORE-GD, which is expected since CORE-Spectral reduces to a single stationary-distribution computation for a Markov transition matrix, whereas CORE-GD solves a smooth convex program by iterative optimization.
Among competing depth-based procedures, CORE-Spectral is typically faster than halfspace, projection, and spatial depths, whose implementations involve repeated optimization or many random projections.
The only faster baselines are Mahalanobis depth and Neg-$\ell_2$, as they require only moment calculations (mean/covariance) and simple Euclidean norms.
Detailed timing results are reported in Supplementary Section~S.3.

\section{Real Data Analysis}
\label{sec:real}

\subsection{Application to the Madelon data}

We study a dataset from the UCI Machine Learning Repository.\footnote{\url{https://archive.ics.uci.edu/dataset/171/madelon}}
The dateset is generated from $32$ clusters positioned on the vertices of a five-dimensional hypercube with $n=2000$ and $p=500$.
Only $20$ features are informative, with the remaining $480$ features serving as non-informative ``probe'' variables.
We compute the score vector $\hat{\btheta}=(\hat\theta_1,\ldots,\hat\theta_n)$  using both CORE-GD and CORE-Spectral.

\begin{figure}
    \centering
    \includegraphics[width=1\linewidth]{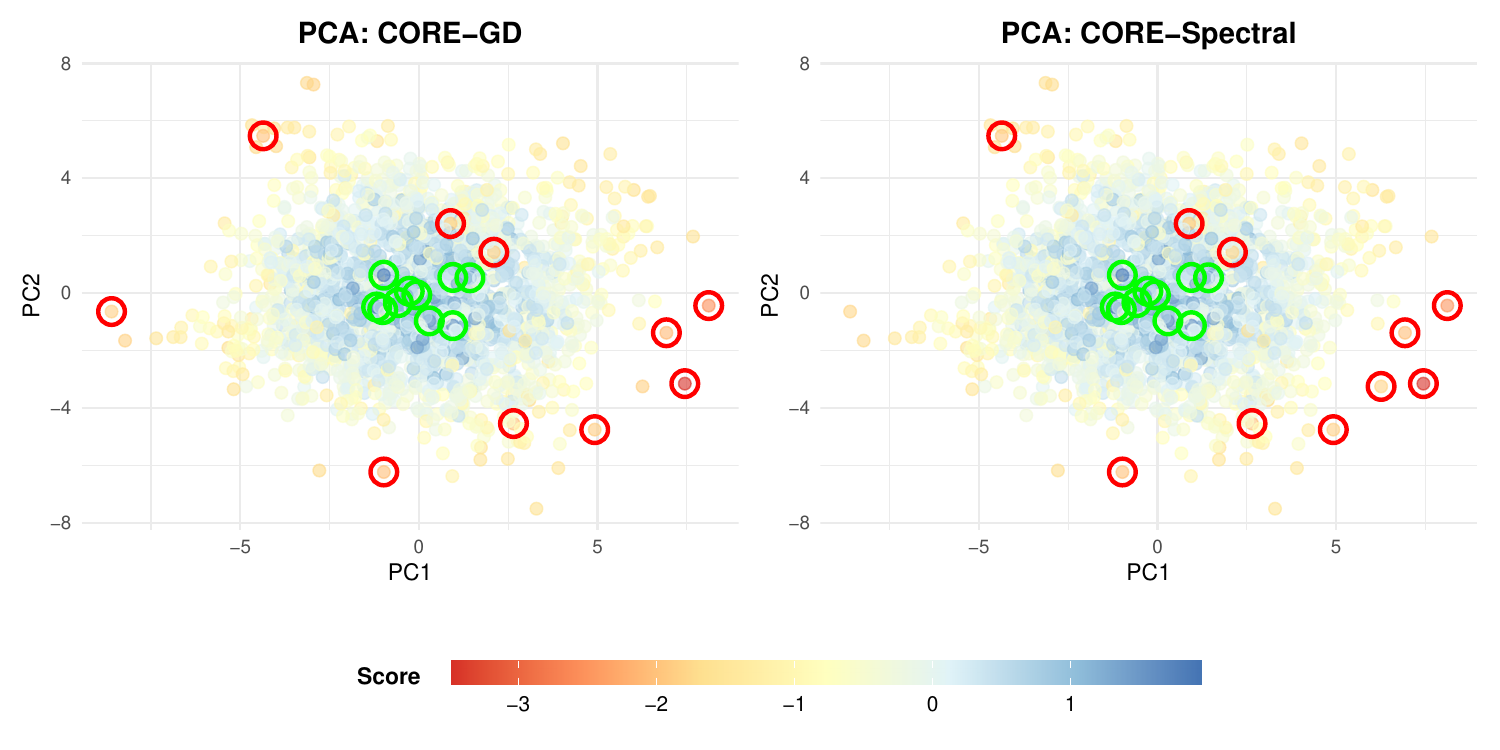}
    \caption{PCA visualization of the Madelon data.
    Left: CORE-GD. 
    Right: CORE-Spectral. 
    Points are colored by the estimated centrality score $\hat\theta_i$ from blue (high) through yellow to orange (low). 
    The $10$ largest-score points are highlighted by green circles and the $10$ smallest-score points are highlighted by red circles.}
    \label{fig:pca}
\end{figure}

Figure~\ref{fig:pca} visualizes the resulting scores on a two-dimensional embedding. 
We apply PCA to the original $500$-dimensional data and project each sample onto the first two principal components. Each point is then colored by its estimated score $\hat\theta_i$, ranging from blue (high) through yellow to orange (low). 
The left and right panels correspond to CORE-GD and CORE-Spectral, respectively. 
In each panel, we mark the $10$ samples with the largest $\hat\theta_i$ values using green circles and the $10$ samples with the smallest $\hat\theta_i$ values using red circles. 
The two panels are nearly identical. 
The highest-score points concentrate near the center of the PCA plane, while the lowest-score points tend to lie toward the periphery. 
This pattern is consistent with the definition of preference-based centrality: points in denser regions of the induced dissimilarity geometry tend to be closer to a typical draw from the distribution and therefore achieve higher values of $r_F(\cdot)$, and hence higher calibrated scores via the monotone link in Theorem~\ref{thm:H-properties}.

\subsection{Application to the Medfly data}

We apply the proposed CORE methodology to the \texttt{medfly25} dataset, which records daily egg-laying counts for female Mediterranean fruit flies (medflies, \emph{Ceratitis capitata}) from the study of \citet{carey1998relationship}. 
Each observation is an egg-laying trajectory observed over the early life span of a fly, and the dataset has been widely used as a benchmark for functional outlier detection.
The preprocessed version provided in the \textbf{fdapace} R package contains $789$ trajectories observed over the first $25$ days after hatch.
To assess stability across repeated experiments, we randomly draw subsamples of size $n=200$ and repeat the analysis over $100$ replicates. 
This subsampling strategy allows us to compute mean and standard deviation of rank correlations across methods.

In the simulation section, we compared CORE with several multivariate depth notions.
However, those depth methods are designed for finite-dimensional Euclidean data and cannot be directly applied to functional trajectories.
Therefore, in this real-data experiment we compare CORE with the regularized halfspace depth (RHD) proposed by \citet{Yeon2025JRSSB}, which is specifically developed for functional data.

For each subsample, we compute three centrality scores: CORE-GD, CORE-Spectral, and RHD.
Pairwise dissimilarities between trajectories are defined using the $L^2$ metric. 
For two trajectories $x_i(t)$ and $x_j(t)$ observed over $\mathcal T$, we set
\[
\delta(x_i, x_j)
=
\left(
\int_{\mathcal T}
\{x_i(t)-x_j(t)\}^2\,dt
\right)^{1/2},
\]
which is approximated numerically using the discretized observations over the first 25 days.

We first compare CORE-GD and CORE-Spectral.
Table~\ref{tab:coregd_vs_corespec_spearman_medfly} reports the Spearman correlation between the two scores over $100$ random subsamples.
The correlation is extremely high ($0.996$ with negligible variability), showing that the spectral approximation yields rankings almost identical to those obtained from the convex M-estimation formulation.
This confirms that CORE-Spectral provides an accurate approximation to CORE-GD.

\begin{table}[ht]
\centering
\caption{Spearman correlation ($\text{Mean}_{\text{SD}}$ over $100$ replicates) between CORE-GD and CORE-Spectral scores on the \texttt{medfly25} dataset.}
\label{tab:coregd_vs_corespec_spearman_medfly}
\small
\setlength{\tabcolsep}{10pt}
\renewcommand{\arraystretch}{1.15}
\begin{tabular}{lc}
\toprule
Comparison & Spearman \\
\midrule
CORE-GD vs CORE-Spectral & $0.996_{0.001}$ \\
\bottomrule
\end{tabular}
\end{table}

\begin{figure}[ht]
    \centering
    \includegraphics[width=1\linewidth]{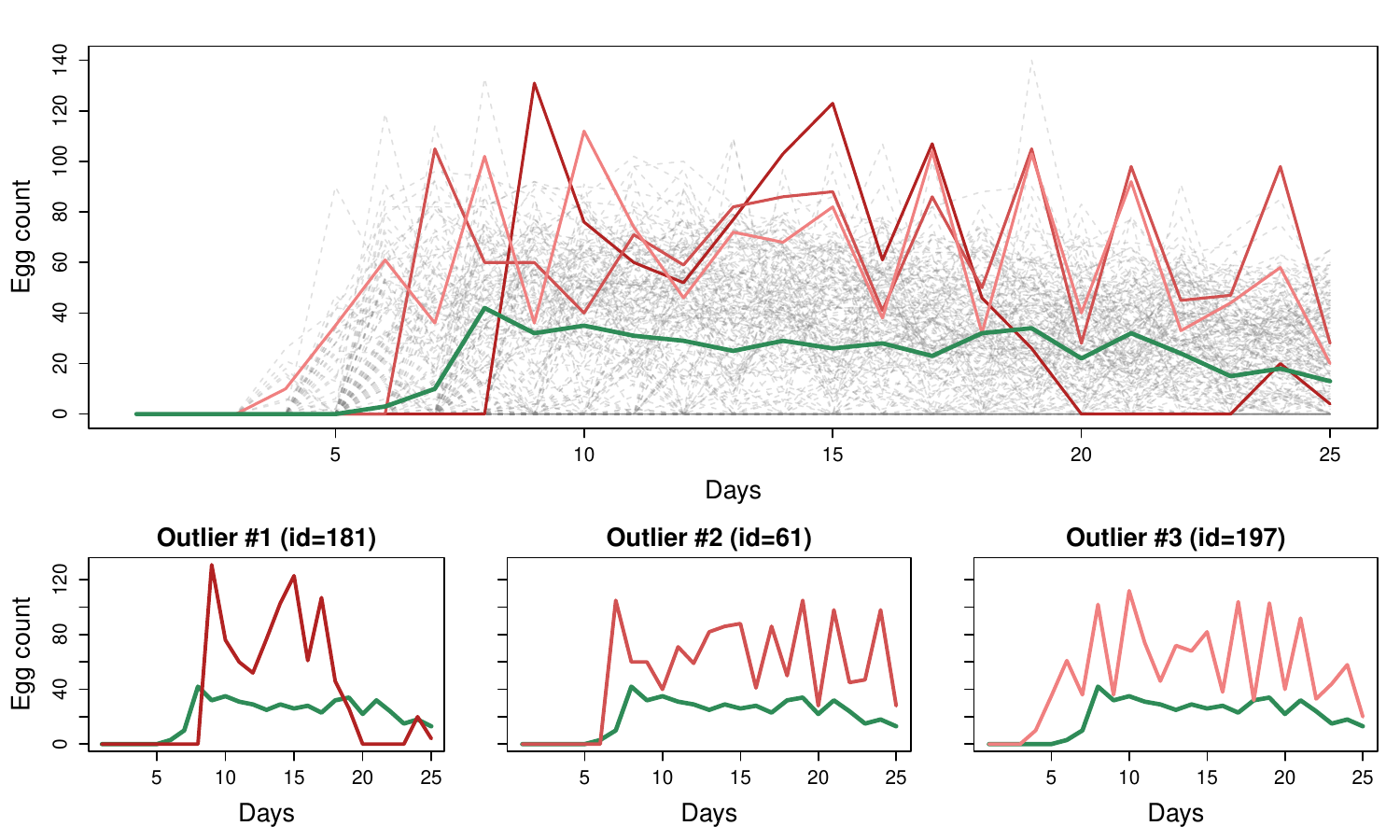}
    \caption{Centrality and outlier visualization of the medfly dataset based on CORE-GD scores.
    Three trajectories (IDs 181, 61, and 197) are detected as outliers by the boxplot rule and highlighted in varying shades of red.
    The top panel displays all curves, with the most central (deepest) trajectory marked in green, while the three detected outliers are separately shown in the bottom panels together with the deepest trajectory.}
    \label{fig:medfly}
\end{figure}

Since the two CORE variants produce nearly identical rankings, we focus on CORE-GD in the remainder of this section.
Figure~\ref{fig:medfly} visualizes the deepest trajectory and detected outliers based on CORE-GD scores for one subsample.
The top panel displays all egg-laying trajectories, with the most central trajectory highlighted in green.
Outliers are identified using the standard boxplot rule with coefficient $1.5$: an observation is flagged if its score falls below the lower fence $Q_1-1.5\times\mathrm{IQR}$, where $\mathrm{IQR}=Q_3-Q_1$.
Using this rule, three trajectories (IDs 181, 61, and 197) are detected as outliers and highlighted in varying shades of red, with darker shades indicating lower centrality.
The bottom panels display each flagged trajectory together with the deepest trajectory to illustrate their deviations in functional shape.

\begin{table}[ht]
\centering
\caption{Spearman correlation ($\text{Mean}_{\text{SD}}$ over $R=100$ replicates) between CORE-GD score and RHD depth on the \texttt{medfly25} dataset, across truncation levels $J$ and quantile levels $u \in \{0.4,0.6,0.8,0.95\}$.}
\label{tab:coregd_vs_rhd_spearman_medfly}
\begin{tabular*}{0.92\linewidth}{@{\extracolsep{\fill}}lcccc}
\toprule
& $u = 0.4$ & $u = 0.6$ & $u = 0.8$ & $u = 0.95$ \\
\midrule
$J=6$  & $0.813_{0.029}$ & $0.780_{0.040}$ & $0.736_{0.053}$ & $0.692_{0.058}$ \\
$J=8$  & $0.803_{0.036}$ & $0.774_{0.045}$ & $0.731_{0.056}$ & $0.689_{0.061}$ \\
$J=10$ & $0.801_{0.035}$ & $0.765_{0.050}$ & $0.726_{0.058}$ & $0.683_{0.063}$ \\
$J=15$ & $0.795_{0.036}$ & $0.758_{0.051}$ & $0.717_{0.060}$ & $0.676_{0.065}$ \\
\bottomrule
\end{tabular*}
\end{table}

Finally, we compare CORE-GD with RHD, which involves two tuning parameters: the truncation level $J$, which controls the number of retained functional components, and the quantile level $u$, which determines the degree of regularization in the depth construction.
To provide a comprehensive comparison, we evaluate RHD over a grid of parameter choices with $J \in \{6,8,10,15\}$ and $u \in \{0.4,0.6,0.8,0.95\}$.
Table~\ref{tab:coregd_vs_rhd_spearman_medfly} reports the Spearman correlation between CORE-GD scores and RHD depths.
The correlations range from $0.676$ to $0.813$, indicating substantial agreement in centrality rankings and supporting the validity of the proposed CORE framework in functional settings.
Higher correlations are observed for smaller values of $J$ and $u$.
Under these settings, RHD emphasizes global magnitude and coarse-scale structure, whereas larger $J$ and $u$ increase sensitivity to local shape features.
Since CORE-GD aggregates global pairwise comparisons under the $L^2$ dissimilarity, its rankings align more closely with the globally oriented RHD configurations.
Taken together, these findings suggest that CORE provides a complementary perspective to functional depth methods, achieving comparable centrality rankings while arising from a fundamentally different pairwise-comparison formulation.

\spacingset{1.45} 

\section{Conclusion and Discussion}
\label{sec:conclusion}

We proposed CORE, a preference-based framework for defining population centrality through reference-based pairwise comparisons on a general dissimilarity space $(\mathcal X,\delta)$. 
Methodologically, we developed two complementary estimators, CORE-GD and CORE-Spectral, which together provide a statistically principled and computationally scalable approach to estimating preference-based rankings.
Our theoretical analysis establishes existence and uniqueness of the population target, consistency of the proposed estimators, and shows that the induced ranking satisfies the canonical properties expected of a statistical depth notion.

Throughout the paper, we focused on the no-tie formulation, which is natural
for continuous distributions where ties occur with probability zero.
When ties are relevant, for example under discrete metrics or quantized
representations, standard tie corrections can be incorporated by assigning
half weight to equality events in the pairwise comparison rule.
Such modifications require only minor technical adjustments and do not alter
the conceptual framework.

Several limitations merit discussion.
First, empirical estimation of pairwise preferences is inevitably affected by
comparison noise and finite-sample bias, which may induce ranking variability
in challenging regimes such as weak signal, heavy tails, or high-dimensional
settings.
Second, the full pairwise construction entails $O(n^2)$ time and memory costs,
which can become prohibitive for very large sample sizes despite the availability
of spectral approximations.

These limitations point to several promising directions for future work.
From a modeling perspective, an important extension is to develop adaptive or
data-driven strategies for selecting or learning the dissimilarity $\delta$ so
that the resulting preference-based centrality aligns more closely with
downstream objectives.
Another direction is to study smoother or probabilistic comparison rules that
reduce sensitivity to small distance perturbations when $\delta$ is noisy or
approximately computed.
From a computational standpoint, scalable variants based on sparse comparison
graphs, pair subsampling, and stochastic or mini-batch optimization schemes
are natural avenues to pursue.
Finally, developing principled uncertainty quantification tools---including
confidence intervals and hypothesis tests for scores and induced rankings---and
exploring richer paired-comparison models beyond the BTL projection remain
important open problems for preference-based centrality and ranking.

\bibliographystyle{chicago}
\bibliography{ref}

\end{document}